\documentclass[12pt,preprint]{aastex}

\shorttitle{Magnetic fields in M dwarfs}
\shortauthors{Phan-Bao et al.}

\begin{document}

\title{Magnetic fields in M dwarfs: rapid magnetic field variability in EV Lac}

\author{Ngoc Phan-Bao,\altaffilmark{1,2} 
Eduardo L. Mart\'{\i}n,\altaffilmark{3,2}
Jean-Fran\c{c}ois Donati,\altaffilmark{4} 
and Jeremy Lim\altaffilmark{1}}

\altaffiltext{1}{Institute of Astronomy and Astrophysics, Academia Sinica, P.O. Box 23-141, Taipei 106, Taiwan, ROC; pbngoc@asiaa.sinica.edu.tw; jlim@asiaa.sinica.edu.tw.}
\altaffiltext{2}{present address (PBN):University of Central Florida, Dept. of Physics, PO Box 162385, Orlando, FL 32816-2385, USA; pngoc@physics.ucf.edu.}
\altaffiltext{3}{Instituto de Astrof\'{\i}sica de Canarias, C/ V\'{\i}a L\'actea s/n, E-38200 La Laguna (Tenerife), Spain; ege@iac.es.}
\altaffiltext{4}{Laboratoire d'Astrophysique, Observatoire Midi-Pyr\'en\'ees, F-31400 Toulouse, France; donati@ast.obs-mip.fr.}

\begin{abstract}
We report here our spectropolarimetric observations obtained 
using the Espadons/CFHT high resolution spectrograph of two M dwarf
stars which standard models suggest are fully convective: EV~Lac (M3.5) and HH~And (M5.5). 
The difference in their rotational velocity makes them good targets to study
the dependence of the magnetic field topology in M dwarfs on rotation. 
Our results reveal some aspects of the field topology in EV~Lac and HH~And.
We measured mean longitudinal magnetic field strengths ($B_{\rm z}$) in EV~Lac
ranging from 18$\pm$3~G to $-$40$\pm$3~G. The $B_{\rm z}$ variations are seen to 
occur in a timescale of less than 50 minutes, significantly shorter than the rotation period, 
and are not due to a flaring event.  
We discuss some formation scenarios of the Zeeman signatures found in EV~Lac. 
For HH~And we could not detect circular polarization and thus we place an upper limit 
to $B_{\rm z}$ of 5~G. 

\end{abstract}


\section{Introduction}

A large number of M dwarfs have been detected in the last few 
decades, and our knowlegde of these faint stars has kept improving, but there are still many  
challenges for research on their magnetic activity.
Main-sequence stars are expected to be fully convective if their mass lies below a certain
value, about 0.3-0.4 $M_{\odot}$ (M3-M4 spectral types) as suggested by the stellar evolution theory;
this probably shifts toward lower masses due to the influence of the magnetic field \citep{cox,mullan}.
In fully convective stars, their magnetic field may be produced by dynamo processes \citep{durney}
that differ from the shell-dynamo at work in partly-convective Sun-like stars. 
However the nature of these dynamos is not clear \citep{liebert}.
Additionally, recent models \citep{dobler,chabrier} disagree with the observations
as pointed out in \citet{donati06}.

One way to understand the nature of magnetic activity in M dwarfs, especially in fully-convective
stars, is to directly measure magnetic fields.
In the two last decades, magnetic fields in late type main sequence stars have been measured for:
G, K dwarf stars \citep{robinson80,marcy84,saar86,basri88,valenti95,ruedi}
or M dwarfs \citep[and references therein]{saar85,johns-krull96,saar01,reiners}. 
The basic idea of the measurement method employed by these authors is a comparison of differential 
Zeeman broadening between magnetically sensitive
and insensitive spectral lines, through unpolarised high-resolution spectra. 
These line broadening measurements have indicated that in some active M dwarfs the 
stellar surface is covered with magnetic fields of 3-4~kG and the filling
factor of about 50\%; however they are poorly informative on the field topology.
One should note that an early attempt to search for circular polarisation in M dwarfs
was carried out by \citet{vogt} but there was no clear detection.
In this paper, we report our analysis of polarised high-resolution spectra of 2 M-dwarf stars: 
a rapidly rotating M3.5 (EV~Lac) and a slowly rotating M5.5 (HH~And). 
Based on their spectropolarimetry presented in Sec.~2, we discuss some interesting aspects on 
the complex field topology in EV~Lac and HH~And in Sec.~3.
\section{Targets, spectropolarimetric observation and data reduction}
\subsection{Targets}
\subsubsection{EV Lac (Gl 873)}
The star is a fast rotator for an M dwarf: $vsini = 4.5$~km/s \citep{johns-krull96},
with this rotational velocity EV~Lac is expected to have a strong magnetic field.
This is therefore a good target to study the dependence of magnetic
dynamo on rapidly rotating M stars.
EV~Lac is also well-known as an M3.5 flare star \citep[and references therein]{osten};
the strong $H_{\alpha}$ emission observed \citep{stauffer} indicates a high chromospheric
activity level in EV~Lac. The rotational period of the star is about 4 days, derived from its rotational velocity
and a corresponding radius of an M3.5 dwarf ($R \sim 0.36R_{\odot}$, \citealt{chabrier97, favata}).
\citet{pettersen92} have also well determined a photometric period of 4.4~days for EV~Lac. 
The mean field measurements 
using synthetic spectrum fitting have previously been reported:
$|B|$ = 4.3 kG, $f = 85$\% 
in \citet{saar94b}; and
$|B|$ = 3.8 kG, $f=50$\% in \citet{johns-krull96}.
However, no measurements of its longitudinal magnetic field $B_{\rm z}$
have been reported so far. 
\subsubsection{HH And (Gl 905)}
The M5.5 dwarf star in contrast with EV~Lac is a slow rotator: $vsini < 1.2$~km/s \citep{delfosse98}.
This is therefore a good target to study the dependence of magnetic dynamo on slowly rotating M stars.
No $H_{\alpha}$ emission detected \citep{stauffer} indicates a very low chromospheric activity level
in HH~And.
Its rotational period is longer than 7 days, derived from an upper limit for $vsini$ as given above
and an M5.5 dwarf radius ($R \sim 0.17R_{\odot}$, \citealt{chabrier97}).
There are no measurements of its magnetic field up to now.
The X-ray observations \citep{schmitt} of nearby stars have implied
that the coronal activity level in EV~Lac is from 10 to 100 times higher than in HH~And.
\subsection{Spectropolarimetric observation and data reduction}
We observed EV~Lac and HH~And with the Espadons/CFHT high resolution
spectrograph ($R = 65,000$; \citealt{donati03}) providing a wavelength coverage of 370-1,000~nm, 
in spectropolarimetric mode to measure Stokes $I$ and $V$ parameters. 
For each star we took three successive exposures, 50 minutes of each
exposure for EV~Lac and 40 minutes for HH~And. 
We obtained high signal-to-noise spectra 
of about 270:1 and 150:1 for EV~Lac and HH~And, respectively.
Wavelength calibrated unpolarised and polarised spectra corresponding to each
observing sequences are extracted with the dedicated software package Libre-ESpRIT 
(\citealt{donati97}; Donati et al., in preparation)
following the principles of optimal extraction \citet{horne}.

To compute the mean longitudinal magnetic field of the stars from the photospheric lines, we use
the Least-Squares Deconvolution (LSD) multi-line analysis procedure\footnote{
For this work, we used a line list \citep{kurucz} corresponding to M spectral types
matching that of EV~Lac and HH~And.
About 5,000 intermediate to strong atomic spectral lines with the average Land\'e factor $g_{\rm eff}$
of about 1.2 are used simultaneously to retrieve the average polarisation information in line profiles, with typical
noise levels of $\approx$0.06\% (relative to the unpolarised continuum level) per
1.8~km/s velocity bin and per individual polarisation spectrum, corresponding
to a multiplex gain in S/N of about 10 with respect to a single average line analysis.
}
given in \citet{donati97}, producing mean Stokes $I$ (unpolarised) and $V$ (circularly polarised)
profiles for all collected spectra.
Figures 1 and 2 show our mean Stokes $V$ profiles of EV~Lac and HH~And. 
In the case of EV~Lac, 
mean longitudinal field strengths corresponding to three
successive exposures are: 
$B_{\rm z}$ = 18$\pm$3~G, $-$40$\pm$3~G and
$-$37$\pm$3~G, respectively.
It is very interesting to note that using the formula given in \citet{wade}
the mean longitudinal field strengths estimated from the
chromospheric lines are much stronger than from the photospheric lines;
and they are different between the Balmer lines ($B_{\rm z} = 260\pm10$~G) and the Ca~II infra-red triplet (IRT)
($B_{\rm z} = 150\pm20$~G, an average over the three triplet lines). We defer a discussion of this result to Sec.~3.

For HH~And, its Stokes $V$ profiles from our observations have not revealed a significant magnetic field, 
$B_{\rm z} = -5\pm2$~G. We have therefore set up an upper limit of 5~G on $B_{\rm z}$ for the star.
\section{Discussion}
In this paper, we could not detect the circularly polarised signature in HH~And
but we did detect it in EV~Lac. 
The explanation might be a difference in the kind of magnetic dynamo that
dominates in each star.
Since EV~Lac (M3.5) and HH~And (M5.5) have low enough masses, 
$\sim 0.35 M_{\odot}$ for EV~Lac and $\sim 0.15 M_{\odot}$ for HH~And \citep{delfosse00}, 
and they are therefore expected to be
fully convective \citep{chabrier97}. If this is the case, a turbulent
dynamo \citep{durney} may dominate in HH~And and generates a small-scale structure
that is not accessible to us.
In the case of EV~Lac, the magnetic field is probably generated from
both a turbulent dynamo and another dynamo kind.
The latter may only work in rapidly rotating stars
and produces a large-scale structure
whose magnetic fields are able to be detected, 
e.g. EV~Lac (this paper) or V374~Peg \citep{donati06}.

Figures 3-6 show the Stokes $V$ and $I$ profiles of the H$_{\alpha}$ at 656.28~nm, and the Ca~II~IRT
(at 849.802, 854.209, and 866.214~nm) respectively for the last exposure.
The difference in the Stokes $V$ profile between the H$_{\alpha}$ 
and the Ca~II~IRT reflects: (1) an imhomogeneous and complex field topology in EV~Lac;
(2) the different shapes of the emission cores ($V \sim \delta I/\delta \lambda$);
(3) the difference in the effective Land\'e factor and the $\sigma$ component distribution
of the lines; (4) the difference in line formation heights of the H$_{\alpha}$ and the Ca~II~IRT,
and the field strengths at those heights \citep{vernazza}.

During our observation, the Balmer lines as well as the emission cores
of the Ca~II~IRT in EV~Lac indicate chromospheric (and therefore magnetic) activity.
This supports the idea that the Zeeman signatures of the Balmer lines and the Ca~II~IRT have 
a chromospheric origin. 
On the other hand, the longitudinal magnetic field strength computed 
from the Balmer lines ($B_{\rm z} = 260$~G) is much stronger 
than that from the Ca~II~IRT ($B_{\rm z} = 150$~G),
implying that the Zeeman signatures of the Balmer lines and the Ca~II~IRT might be 
formed at different heights \citep{vernazza}.

Our observations have also indicated a significant variability of the field in EV~Lac:
an opposite sign of $B_{\rm z}$ observed in the first exposure compared with the second and the third one.
The time scale of this variability is very short, i.e. less than 50 minutes.
We consider several possibilities that could explain the variability.
First, the possibility of star rotation changing the field vector is ruled out
since in the case of EV~Lac its rotational period of 4 days could not significantly change 
the field vector 
in 50 minutes or $\sim$0.8\% of a whole phase.
Second, the possibility of a strong flare during our observations,  
modifying the Zeeman signature obtained in the first exposure, 
is also precluded since we do not find
any significant change in the H$_{\alpha}$ profile whose equivalent width is of about 
3.3~\AA~for all three exposures.
Finally, with high probability the magnetic field in EV~Lac is intrinsically variable.
This again indicates the imhomogeneous and complex field in the star
that probably leads observational results as discussed above.

More spectropolarimetric observations at different rotational phases 
will allow us to reconstruct the EV~Lac field topology
not only on the stellar surface but also in three dimensions by using different lines coming
from different layers of the star. 
This will lead to a strong improvement of our understanding on magnetic field
in EV~Lac and hence in low mass stars.

\acknowledgments
We thank the referee for many helpful comments that clarified and significantly improved our paper.
P.B.N. acknowledges useful discussions with J. Valenti and G. Chabrier. 
This paper is based on observations obtained at the Canada-France-Hawaii Telescope (CFHT)
which is operated by the National Research Council of Canada, the Institute National des Sciences 
de l'Univers of the Centre National de la Recherche Scientifique of France, and the University of Hawaii.

\clearpage

\begin{figure}
\vskip 1in
\hskip -0.25in
\centerline{\includegraphics[width=4.5in,angle=-90]{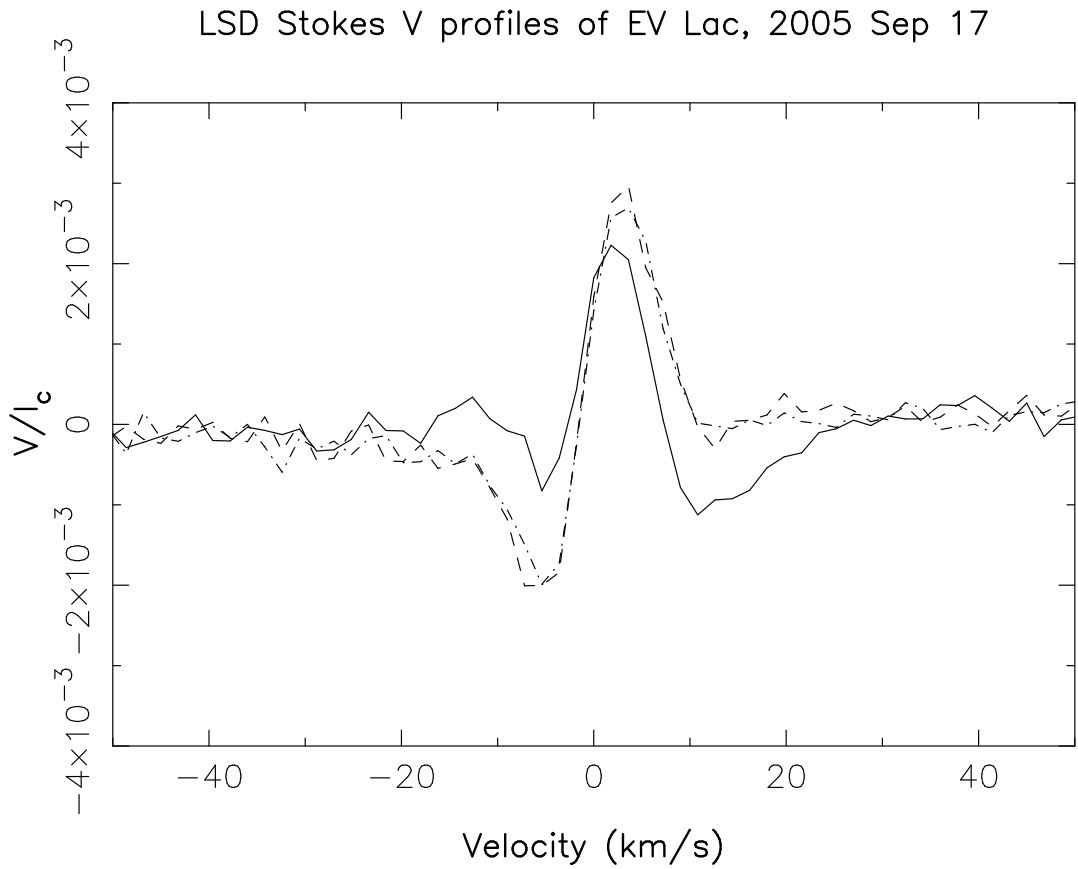}}
\caption{Stokes $V$ profiles of EV Lac (Gl 873, M4.5); solid line: first
exposure, dashed line: second exposure, and dash-dotted line: third exposure; 50 minutes
for each exposure were taken. A positive blueward component (solid line) implies
a mixed polarity in the first exposure.
\label{StokesV_EVLac}}
\end{figure}

\clearpage

\begin{figure}
\vskip 1in
\centerline{\includegraphics[width=4.5in,angle=-90]{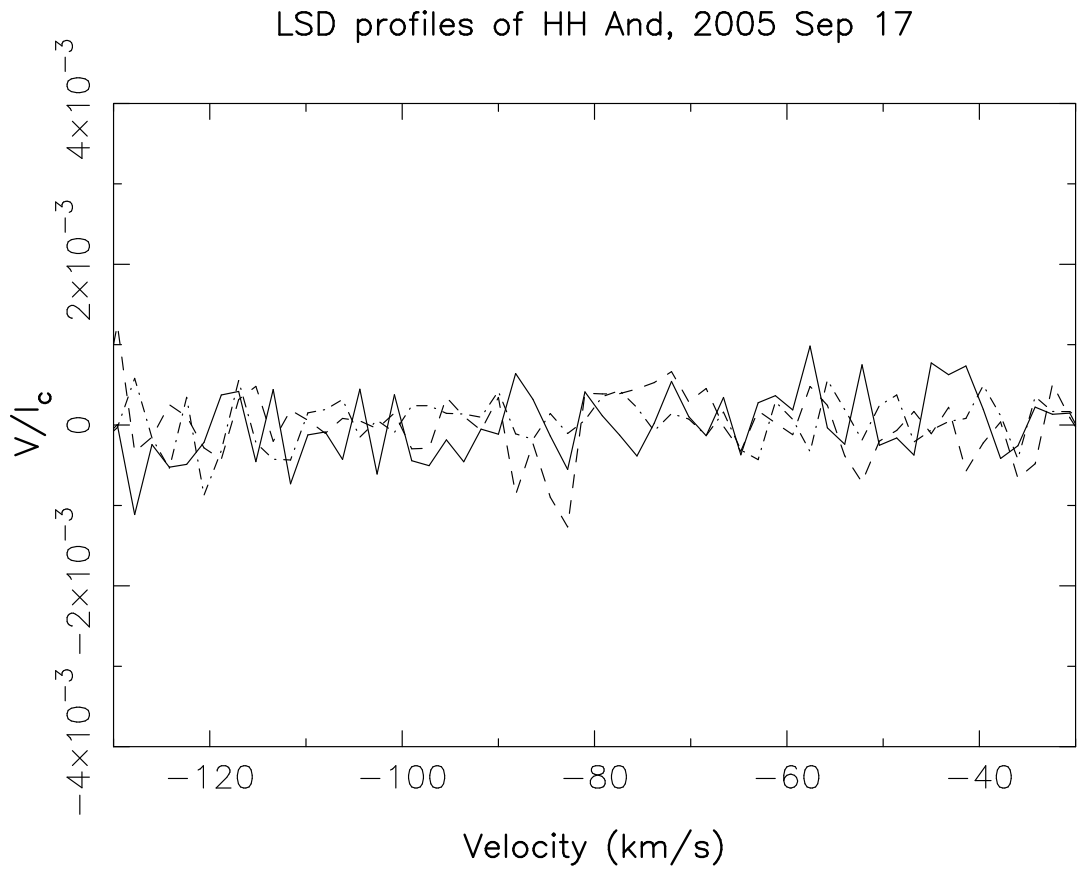}}
\caption{Stokes $V$ profiles of HH And (Gl 905, M5); three exposures were also
taken, 40 minutes for each one. 
No longitudinal magnetic field has been detected, we set up an upper limit of 5~G on $B_{\rm z}$.
\label{StokesV_EVLac}}
\end{figure}

\clearpage

\begin{figure}
\centerline{\includegraphics[width=4.5in,angle=-90]{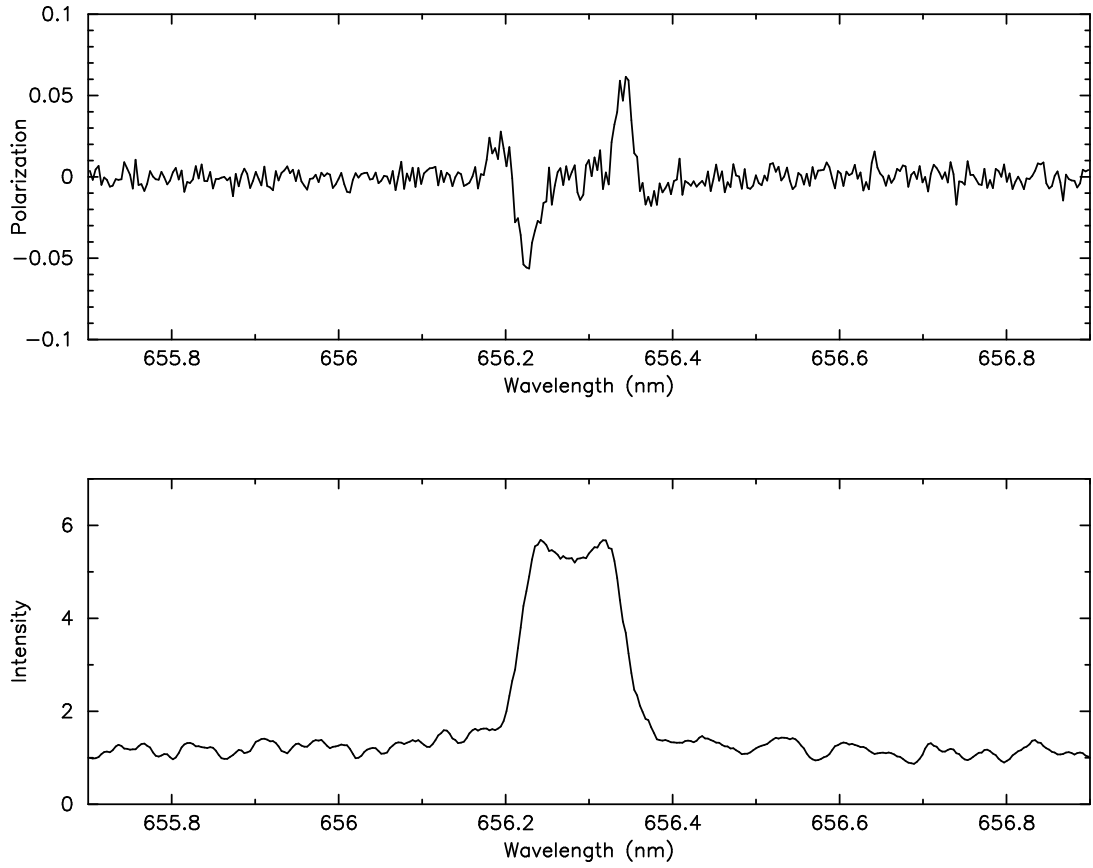}}
\caption{Stokes $V$ (top) and $I$ (bottom) profile  (exposure \#3) of the H$_{\alpha}$ line at 656.28~nm.
Additional components in the Stokes $V$ profile (a positive component
at 656.19~nm and a negative one at 656.38~nm) implies a mixed polarity. 
\label{Halpha}}
\end{figure}

\clearpage

\begin{figure}
\centerline{\includegraphics[width=4.5in,angle=-90]{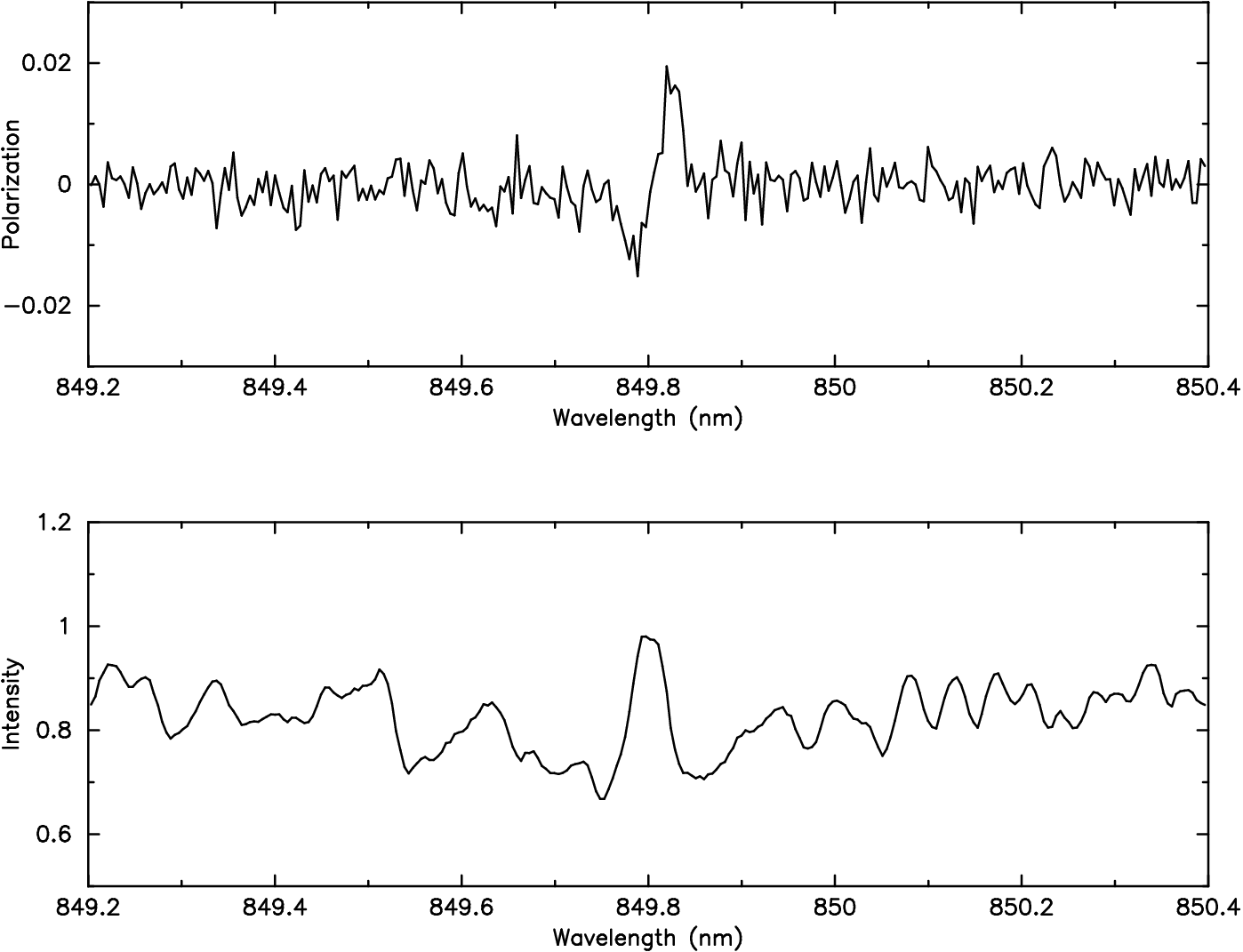}}   
\caption{Stokes $V$ (top) and $I$ (bottom) profile (exposure \#3) of the
Ca~II IRT at 849.802~nm.
\label{CaIIIRT1}}
\end{figure}

\clearpage

\begin{figure}
\centerline{\includegraphics[width=4.5in,angle=-90]{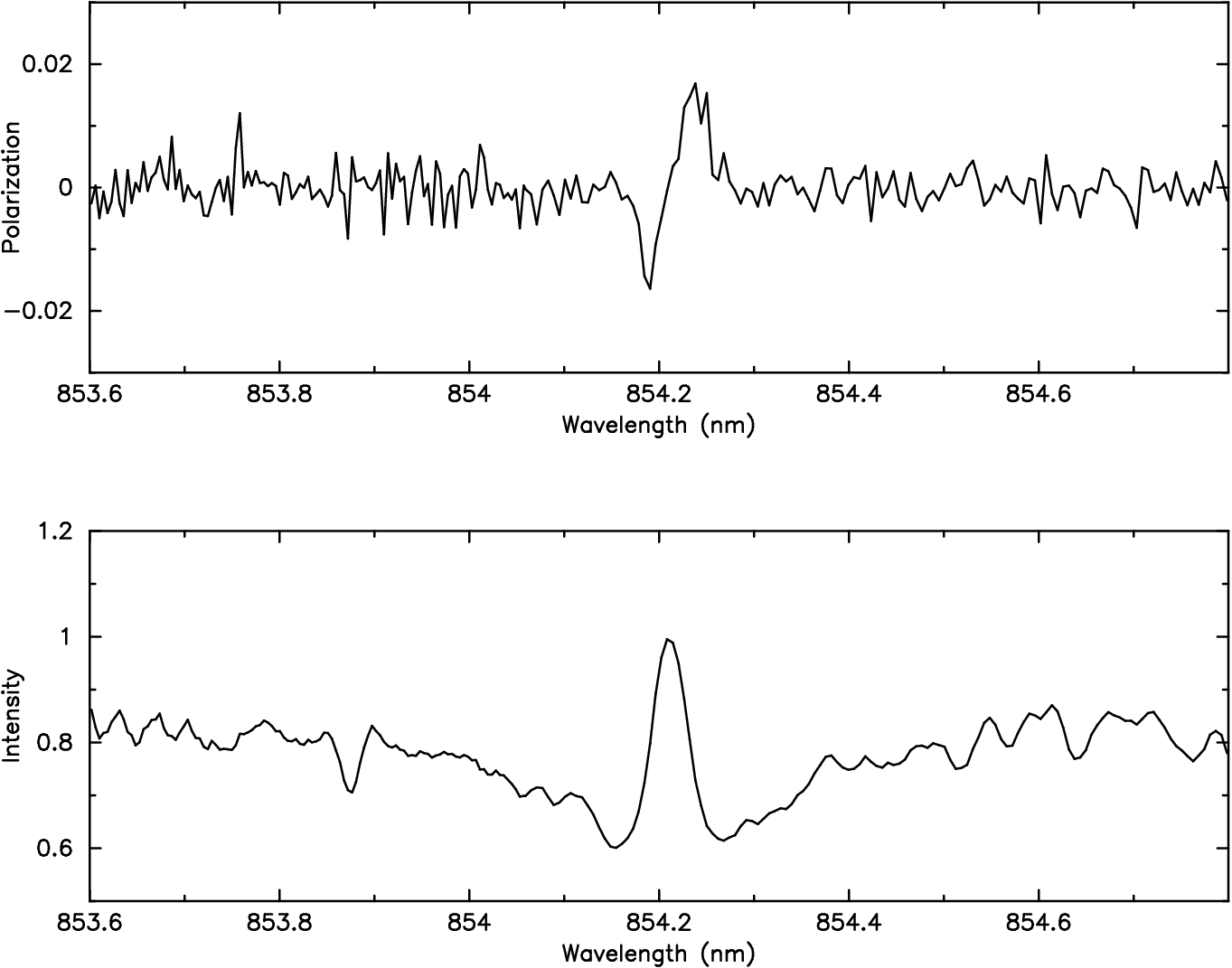}}
\caption{Stokes $V$ (top) and $I$ (bottom) profile (exposure \#3) of the Ca~II IRT at
854.209~nm.
\label{CaIIIRT2}}
\end{figure}

\clearpage

\begin{figure}
\centerline{\includegraphics[width=4.5in,angle=-90]{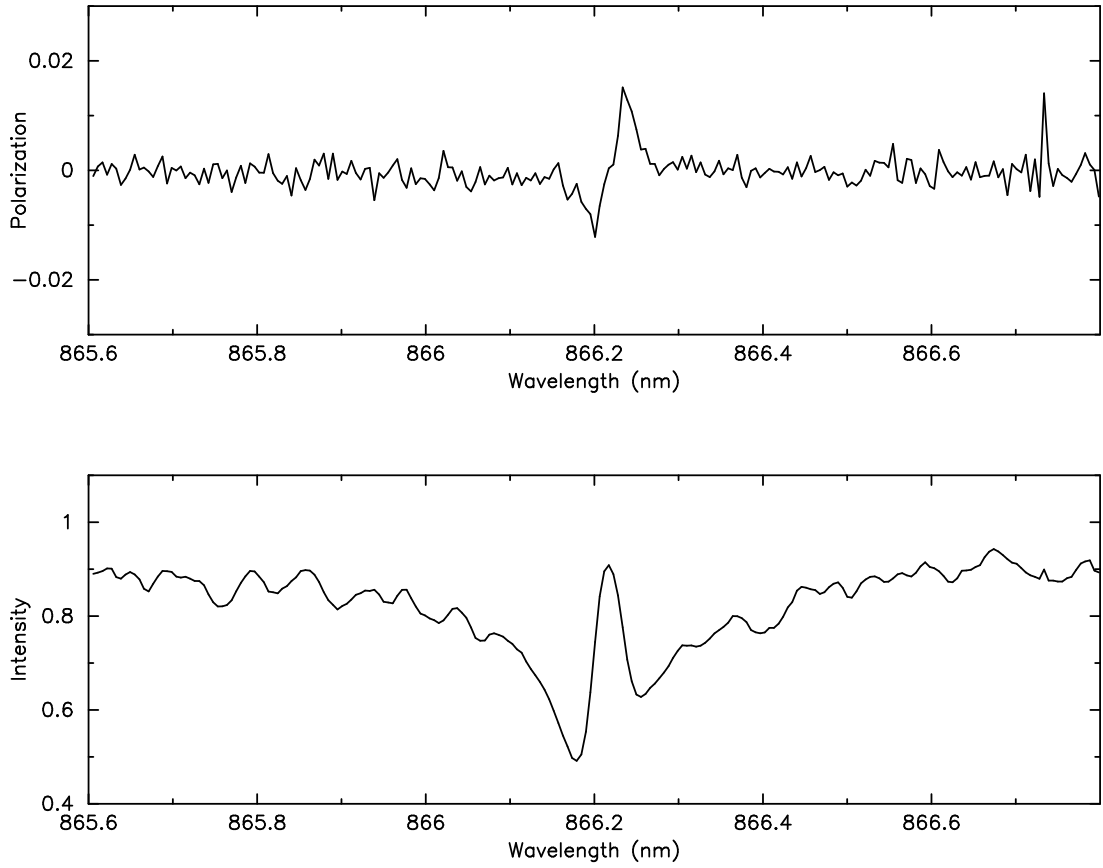}}
\caption{Stokes $V$ (top) and $I$ (bottom) profile (exposure \#3) of the Ca~II IRT at
866.214~nm.
\label{CaIIIRT3}}
\end{figure}


\begin{thebibliography}{}

\bibitem[Basri \& Marcy(1988)]{basri88}
Basri, G., \& Marcy, G. W. 1988, \apj, 330, 274

\bibitem[Chabrier \& Baraffe(1997)]{chabrier97}
Chabrier, G., \& Baraffe, I. 1997, \aap, 327, 1039

\bibitem[Chabrier \& K\"uker(2006)]{chabrier}
Chabrier, G., \& K\"uker, M. 2006, \aap, 446, 1027

\bibitem[Cox et al.(1981)]{cox}
Cox, A. N., Shaviv, G., \& Hodson, S. W. 1981, \apj, 245, L37

\bibitem[Delfosse et al.(1998)]{delfosse98}
Delfosse, X., Forveille, T., Perrier, C., \& Mayor, M. 1998, \aap, 331, 581

\bibitem[Delfosse et al.(2000)]{delfosse00}
Delfosse, X., Forveille, T., S\'egransan, D., et al. 2000, \aap, 364, 217

\bibitem[Dobler et al.(2006)]{dobler}
Dobler, W., Stix, M., \& Brandenburg, A. 2006, \apj, 638, 336

\bibitem[Donati et al.(1997)]{donati97}
Donati, J.-F., Semel. M., Carter, B. D., et al. 1997, \mnras, 291, 658

\bibitem[Donati(2003)]{donati03}
Donati, J.-F. 2003, ASP Conference Proceedings, 307, 41

\bibitem[Donati et al.(2006)]{donati06}
Donati, J.-F., Forveille, T., Cameron, A. C., et al. 2006, Science, 311, 633

\bibitem[Durney et al.(1993)]{durney}
Durney, B. R., De Young, D. S., \& Roxburgh, I. W. 1993, Sol. Phys., 145, 207

\bibitem[Favata et al.(2000)]{favata}
Favata, F., Reale, F., Micela, G., Sciortino, S., et al. 2000, \aap, 353, 987

\bibitem[Horne(1986)]{horne}
Horne, K. 1986, PASP, 98, 609

\bibitem[Johns-Krull \& Valenti(1996)]{johns-krull96}
Johns-Krull, C. M., \& Valenti, J. A. 1996, \apj, 459, L95

\bibitem[Kurucz(1993)]{kurucz}
Kurucz, R. L. 1993, CDROM \# 13 (ATLAS9 atmospheric models) and \# 18 (ATLAS9 and SYNTHE routines,
spectral line database).

\bibitem[Liebert et al.(2003)]{liebert}
Liebert, J., Kirkpatrick, J. D., Cruz, K. L., et al. 2003, \aj, 125, 343

\bibitem[Marcy(1984)]{marcy84}
Marcy, G. W. 1984, \apj, 276, 286

\bibitem[Mullan \& MacDonald(2001)]{mullan}
Mullan, D. J., \& MacDonald, J. 2001, \apj, 559, 353

\bibitem[Osten et al.(2005)]{osten}
Osten, R. A., Hawley, S. L., Allred, J. C., et al. 2005, \apj, 621, 398

\bibitem[Pettersen et al.(1992)]{pettersen92}
Pettersen, B. R., Olah, K., \& Sandmann, W. H. 1992, \aaps, 96, 497

\bibitem[Reiners \& Basri(2006)]{reiners}
Reiners, A., \& Basri, G. 2006, arXiv: astro-ph/0602221

\bibitem[Robinson et al.(1980)]{robinson80}
Robinson, R. D., Worden, S. P., \& Harvey, J. W. 1980, \apj, 236,  L155

\bibitem[R\"uedi et al.(1997)]{ruedi}
R\"uedi, I., Solanki, S. K., Mathys, G., \& Saar, S. H. 1997, \aap, 318, 429

\bibitem[Saar \& Linsky(1985)]{saar85}
Saar, S. H., \& Linsky, J. L. 1985, \apj, 299, L47

\bibitem[Saar et al.(1986)]{saar86}
Saar, S. H., Linsky, J. L., \& Beckers, J. M. 1986, \apj, 302, 777

\bibitem[Saar(1994)]{saar94b}
Saar, S. H. 1994, IAU Symp. Infrared Solar Physics, 154, 493

\bibitem[Saar(2001)]{saar01}
Saar, S. H. 2001, ASP Conference Proceedings, 223, 292

\bibitem[Schmitt \& Liefke(2004)]{schmitt}
Schmitt, J.H.M.M, \& Liefke, C. 2004, \aap, 417, 651

\bibitem[Stauffer \& Hartmann(1986)]{stauffer}
Stauffer, J. R., \& Hartmann, L. W. 1986, \apjs, 61, 531

\bibitem[Valenti et al.(1995)]{valenti95}
Valenti, J. A., Marcy, G. W., \& Basri, G. 1995, \apj, 439, 939

\bibitem[Vernazza et al.(1981)]{vernazza}
Vernazza, J. E., Avrett, E. H., \& Loeser, R. 1981, \apjs, 45, 635

\bibitem[Vogt(1980)]{vogt}
Vogt, S. S. 1980, \apj, 240, 567

\bibitem[Wade et al.(2000)]{wade}
Wade, G. A., Donati, J.-F., Landstreet, J. D., \& Shorlin, S. L. S. 2000, MNRAS, 313, 851

\end{thebibliography}
\end{document}